\documentclass[review]{elsarticle}

\usepackage{rotating}
\usepackage{amsfonts}% per usufruire di mathbb{}%
\usepackage{amsmath}
\usepackage{amsthm}
\usepackage{xcolor}
\usepackage{amssymb,amsmath}
\usepackage{subfig}
\usepackage[english]{babel}
\usepackage{graphicx}
\usepackage{tabularx}
\usepackage{enumerate, hyperref}
\usepackage{tikz}
\usetikzlibrary{shapes.geometric,calc}

\theoremstyle{plain}

\theoremstyle{definition}

\theoremstyle{remark}

\begin{document}

\begin{frontmatter}

\title{{\itshape ``Butterfly Effect'' vs Chaos in Energy Futures Markets.}}

%% Group authors per affiliation:
\author{Loretta Mastroeni}
\address{Dept. of Economics, University of Roma TRE, via Silvio D'Amico 77, 00145 Rome, Italy.}
\ead{loretta.mastroeni@uniroma3.it}

\author{Pierluigi Vellucci}
\address{Dept. of Basic and Applied Sciences for Engineering, Sapienza University of Rome, Via Antonio Scarpa 16, 00161 Rome, Italy.}
\ead{pierluigi.vellucci@sbai.uniroma1.it}

\begin{abstract}
In this paper we test for the sensitive dependence on initial conditions (the so called ``butterfly effect'') of energy futures time series (heating oil, natural gas), and thus the determinism of those series. This paper is distinguished from previous studies in the following points: first, we reread existent works in the literature on energy markets, enlightening the role of \emph{butterfly effect} in chaos definition (introduced by Devaney), using this definition to prevent us from misleading results about ostensible chaoticity of the price series. Second, we test for the time series for sensitive dependence on initial conditions, introducing a coefficient that describes the determinism rate of the series and that represents its reliability level (in percentage). The introduction of this reliability level is motivated by the fact that time series generated from stochastic systems also might show sensitive dependence on initial conditions. According to this perspective, the maximum reliability level obtained here is too low to be able to ensure that there is strong evidence of sensitive The maximum reliability level obtained here was been $\simeq 56\% $, too low to ensure strong evidence of sensitive dependence on initial conditions.

%We point out that the butterfly effect is not equivalent to chaos, although some researchers assume it, but it is a prerequisite for chaoticity. We reread existent works in the literature on energy markets enlightening the role of sensitive dependence on initial conditions in chaos definition. The mathematical definition of chaos recalled here (introduced by Devaney) is helpful to prevent us from misleading results about ostensible chaoticity of the price series. Besides, we test for the time series for sensitive dependence on initial conditions introducing a coefficient $\kappa$ that describes the \emph{determinism rate} of the series and that represents its reliability level (in percentage). The introduction of this reliability level is motivated by the fact that time series generated from stochastic systems also might show sensitive dependence on initial conditions. According to this perspective, the maximum reliability level obtained here is too low to be able to ensure that there is strong evidence of sensitive dependence. \textcolor[rgb]{1.00,0.00,0.00}{introdurre risultati}

JEL classification: C45; C53; D40; Q47.
\end{abstract}
%After making a review of previous results on chaos in energy futures time series, we show a re-reading of them according to a rigorous definition of Chaos, the Devaney's Chaos definition. We analyze the determinism rate $\kappa$ of each time series considered here. Afterward, to check sensitive dependence we estimate the Largest Lyapunov exponents. Since time series generated from stochastic systems also might show positive maximal Lyapunov exponents, we introduce a reliability level (the rate $\kappa$) associated to maximal Lyapunov exponents results. This rate tells us how much the results on sensitive dependence can be considered reliable. The maximum reliability level obtained here was been $\simeq 56\% $, too low to be able to ensure that there is strong evidence of sensitive dependence.
%Since a preliminary question is if the treatment of the data can influence the results of the tests (the filtered data may give a false indication of chaos), in our approach we use only the original data and we leave the question of the filtering to future works.

\begin{keyword}
nonlinear dynamics, chaos, butterfly effect, energy futures
\end{keyword}

\end{frontmatter}

%\linenumbers

\section{Introduction}
\label{sec:intro}

In recent decades, chaos theory has been considered in various scientific fields such as economics, finance, physics, and much others. An important reason to be interested in chaotic behaviour is that it can potentially explain fluctuations in many time series which appear to be random. In particular, as for financial markets, evidence on deterministic chaos would have important implications for regulators and short-term trading strategies. The question is whether such random-looking data is really random or it is completely deterministic. If it is completely random, then its behaviour is not predictable anyway; otherwise, it is possible to predict a deterministic system on short periods of time (instead, long prediction is impossible, due to instability of chaotic systems). Hence, this distinction provides the predictability degree of the analyzed system.

%A very interesting question is to know whether such random-looking data is really random or it is completely deterministic. If it is completely random, then its behaviour is not predictable anyway; otherwise, it is possible to predict a deterministic system in the short periods of time (instead, long prediction is impossible, due to instability of chaotic systems). Hence, this distinction provides the predictability degree of the analyzed system.

Chwee \cite{chwee1998chaos} tests for the presence of chaos using the NYMEX 1-month, 2-month, 3-month, and 6-month daily natural gas settlement prices, from April 1990 to September 1996. In doing so, he uses the BDS statistics and the Lyapunov spectra to determine to what degree futures data resemble a chaotic system. The results fail to provide significant evidence of deterministic chaos. %``Similar causes have similar effects'' is invalid in chaotic systems except for short periods.
Serletis and Gogas \cite{serletis1999north} test for deterministic chaos in seven Mont Belview, Texas hydrocarbon markets, using monthly data from 1985:1 to 1996:12 (the markets are those of ethane, propane, normal butane, iso-butane, naptha, crude oil, and natural gas). In their paper, they estimate the largest Lyapunov exponent, finding an evidence of chaotic process.
Panas and Ninni \cite{Panas00} investigate chaotic structure in daily price data for two major petroleum markets, namely those of Rotterdam and the Mediterranean. The sample consists of the daily prices of different oil products from 4 January 1994 to 7 August 1998, resulting in 1161 observations. All prices were collected from OPEC. The main empirical results obtained by Panas and Ninni's analysis are summarised in Table 5 on \cite{Panas00}. The criteria and methods used here are: correlation dimension; entropy; maximal Lyapunov exponent; Eckmann-Ruelle condition; Brocks or residual test theorem; BDS statistic test. They show ``strong evidence of chaos in a number of oil products considered''. Adrangi et al. \cite{Adrangi01} investigated the presence of low-dimensional chaotic structure in crude oil, heating oil, and unleaded gasoline futures prices from the early 1980s. Daily returns data from the nearby contracts are diagnosed employing correlation dimension test, the BDS test and Kolmogrov entropy. While they find ``strong evidence of non-linear dependence in the data, the evidence is not consistent with chaos''. Moshiri and Foroutan \cite{moshiri2006forecasting} examine daily crude oil futures prices from 1983 to 2003, listed in NYMEX; they test for chaos embedding dimension, BDS, Lyapunov exponent, and neural networks tests, finding a negative evidence of chaos. Kyrtsou et al. \cite{kyrtsou2009energy}, analyze five energy products (crude oil, gasoline, heating oil, propane, and natural gas) over the period from 1994 to mid-January 2008. They estimate the dominant Lyapunov exponent, which is negative and in every case they reject the null hypothesis of chaotic behavior. Using both metric (correlation dimension and Lyapunov exponents) and topological methodologies (recurrence plot analysis), Barkoulas et al. \cite{barkoulas2012metric} consider a data set which consists of daily oil spot prices covering the period 1.2.1985 - 8.31.2011. Applying the ``metric methodologies'', they conclude that both the correlation dimension and the Lyapunov exponents show no chaotic tendencies in the oil market and that ``the test results from both metric and topological methodologies suggest that oil spot prices are the measured footprint of a stochastic rather than a deterministic system'' (pp. 585 \cite{barkoulas2012metric}). Matilla-Garc\'{i}a \cite{matilla2007nonlinear} studies chaotic nature of three energy futures series: natural gas, unleaded gasoline and light crude oil. He investigates the presence of chaos through the largest Lyapunov exponent, finding its ``evidence in futures returns''. In his conclusions, Matilla-Garc\'{i}a writes: ``A natural question arises: Does evidence of chaos depend on the test procedure used by the researcher? This question is left for future research''.

%For an overview of the results of the above mentioned papers, in the framework of a rigorous definition of chaos, see Conclusions in Section \ref{sec:exp}.

Broadly speaking, one uses the term ``chaos'' as a synonym of the sensitive dependence on initial conditions, which is the most popular property of a chaotic system, maybe for its intuitive meaning: tiny differences become amplified. The shorthand is the ``butterfly effect'', a term that has inspired novels and movies. (Let's remember, for example, Ian Malcolm, the mathematician of Jurassic Park.).
This property has been introduced for the first time in a formal definition of chaos by Devaney \cite{Devaney89} \footnote{In literature, there are other definitions of chaos, as for example \cite{Li75}, \cite{Touhey97}, \cite{Martelli98}, \cite{Wiggins13}. Some of them are equivalent to Devaney's one, some others are, generally, not easily applicable.}. According to Devaney, the ingredients of chaos are: sensitive dependence on initial conditions; topological transitivity; density of periodic points.
Nevertheless, all the definitions of ``chaos'' in economic and financial literature, in particular all the papers that we have cited above (\cite{chwee1998chaos}, \cite{serletis1999north}, \cite{Panas00}, \cite{Adrangi01}, \cite{moshiri2006forecasting}, \cite{matilla2007nonlinear}, \cite{kyrtsou2009energy}; see \cite{barkoulas2012metric} for the ``metric methodologies''), refer only to the first property introduced by Devaney, which represents, in some way, an ``experimental'' definition of chaos. Even though it can be checked numerically, it is not equivalent to chaos and might be misleading (see also \cite{yousefpoor2008looking} and the counterexample 3.3 in \cite{Martelli98}). Besides, it can be shown that the butterfly effect is redundant with the other two conditions of Devaney's definition, meaning that butterfly effect is implied by the other two conditions. All these questions are not negligible since also some time series generated from stochastic systems might show sensitive dependence on initial conditions \cite{tanaka1996lyapunov}, \cite{ikeguchi1997lyapunov}, \cite{tanaka1998analysis}.

According to butterfly effect, if two initial conditions have a small difference $\delta x$, their difference after time $\delta t$ will be $\delta x e^{\lambda \delta t}$ with $\lambda > 0$, that is, with exponential separation. Small differences in initial conditions (such as those due to rounding errors in numerical computation) yield widely diverging outcomes for such dynamical systems, making long-term prediction impossible in general.
Lyapunov exponents determine the rate of divergence or convergence of initially nearby trajectories in the phase space \cite{Ko05}, \cite{Ott93}, \cite{Schu89}, \cite{Stro94}. The most important observation is that the largest Lyapunov exponent, denoted as $\lambda_{max}$, uniquely determines whether the system shows butterfly effect or not. If $\lambda_{max}>0$ two initially nearby trajectories of the attractor diverge exponentially fast as time progresses, showing a butterfly effect. Lyapunov exponents represent the rate at which the system creates or distorts information. In fact, let us consider two time series, which have different largest Lyapunov exponents, respectively $\lambda_{max}^1>0$ and $\lambda_{max}^2>0$. If $\lambda_{max}^1/\lambda_{max}^2>1$, then the greater this rate, the faster the propagation of the error in the first time series (compared to the second one).

The aim of this paper is twofold. First, we reread existent works in the literature of energy markets enlightening the role of sensitive dependence on initial conditions in chaos definition: the mathematical definition of chaos recalled here (according to Devaney \cite{Devaney89}) is helpful to prevent us from misleading results about ostensible chaoticity of the returns series. Second, we test for sensitive dependence on initial conditions introducing a coefficient $\kappa$ that describes the \emph{determinism rate} of the analyzed time series, that represents, in percentage, its reliability level. The introduction of this reliability level is motivated by the fact that, as we have already said, time series generated from stochastic systems might show sensitive dependence on initial conditions.

%In the Conclusions of \cite{matilla2007nonlinear}, Matilla-Garc\'{i}a asks: ``A natural question arises: Does evidence of chaos depend on the test procedure used by the researcher? This question is left for future research''.
We can state, according to the above considerations, also in order to give an answer to Matilla-Garc\'{i}a's conclusions, that experimental results should be independent from the test used to obtain them: the basic question is whether the treatment of the data can influence the results of the tests.
Chaos tests can be conducted on both raw as well as filtered data. Empirical analysis of identification of chaotic structure of time series necessarily raises the question of filtering (the transformation of raw data prior to its analysis), because the filtered data may give a false indication of chaos. In \cite{Panas00}, Panas and Ninni compare filtered and raw data, pointing out that the obtained results on the butterfly effect are approximatively the same, even if, generally, filtering may affect the dimensionality of the original data (see \cite{chen1993searching}, as reported also in \cite{Panas00}) and the filtered data may mimic a chaotic behaviour. However, in our approach we have used only raw data and we have left the question of filtering to future works.

%The paper is organized as follows. In Section \ref{sec:intro2} we consider the implications of chaos in energy futures markets, and we introduce one of the most popular definition of chaos, the Devaney's definition, suitable to our purposes. Section \ref{sec:prel} is devoted to the mathematical methodologies of the tests we conducted here, in comparison with some others that appear in the literature. In Section \ref{sec:exp} we explain the results of the paper. Section \ref{sec:conc} is devoted to the conclusions while the Appendix (Section \ref{sec:app}) contains all the tables concerning the numerical evaluations.

The paper is organized as follows. In Section \ref{sec:intro2} we consider the implications of chaos in energy futures markets, and we introduce one of the most popular definition of chaos, the Devaney's definition, suitable to our purposes. In Section \ref{sec:exp} we explain the results of the paper. Section \ref{sec:conc} is devoted to the conclusions while the Appendix 1 (Section \ref{sec:prel}) is devoted to the mathematical methodologies of the tests we conducted here, in comparison with some others that appear in the literature. Appendix 2 (Section \ref{sec:app}) contains all the tables concerning the numerical evaluations.

\section{Implications of ``chaos'' in energy futures time series}
\label{sec:intro2}
In literature (\cite{Devaney89}, \cite{Ko05}, \cite{ruelle1989chaotic}, \cite{Wiggins13}), chaos is considered as an alternative to randomness for systems of ``strange'' behaviour.
One of the most famous (and the simplest) example of chaotic system is the \emph{logistic map} (that underlies the well-known \emph{logistic growth model}); it is written
$$x_{n+1}=r\,x_{n}(1-x_{n})$$
where $x_{n}$ can be represent the ratio of existing population to the maximum possible population and it is a number between zero and one.
The parameter $r>0$ represents the exponential growth rate of the population. The logistic map is \emph{chaotic} (later, we will focus on the meaning of ``chaos'') for $r>2+\sqrt 5 \approx 4.236$ (\cite{Devaney89}, pp. 31 - 50). The quality of unpredictability\footnote{Meaning that chaotic systems are unpredictable in a way that other deterministic systems are not \cite{werndl2009new}.} and apparent randomness led the logistic map equation to be used as a pseudo-random number generator in calculators \cite{gleick2011chaos}. On the other hand, we can say that a chaotic map is a deterministic map which is able to produce ``random looking'' data.

% A very interesting question is to know whether such random-looking data is really random or it is completely deterministic. If it is completely random, then its behaviour is not predictable anyway; otherwise, it is possible to predict a deterministic system in the short periods of time (instead, long prediction is impossible due to instability of chaotic systems). Hence, this distinction provides the predictability degree of the analyzed system.

Another notable system that has chaotic solutions for certain parameter values and initial conditions and that represents a starting point in the whole literature on chaos, is the \emph{Lorenz system} \cite{Lo63}, a simplified mathematical model for atmospheric convection.

From a technical standpoint, in both cases we start from a mathematical model (one or more equations).

Nevertheless, in many practical situations (such as in financial markets and, for our aims, in energy markets) we may not have available a mathematical model and could be necessary to work on time series. Generally, the accurate structural modelling of commodities, and in particular, of the energetic commodities, could be considered impossible. Therefore, evidence of chaos could offer some strategies for modelling price behaviour by simply employing the time series of prices.

%There are many disparate domestic and international factors that affect the prices of these contracts, but the quantifiable information on such factors is generally incomplete.

There are many possible definitions of chaos, ranging from measure theoretic notions of randomness in ergodic theory to the topological approach. In our paper we will focus on the Devaney's definition of chaos \cite{Devaney89} and we refer to working paper \cite{MasVel}, for a mathematical detailed study of the theoretical background (only mentioned here).

Devaney wrote: ``To summarize, a chaotic map possesses three ingredients: unpredictability, indecomposability, and an element of regularity. A chaotic system is unpredictable because of the sensitive dependence on initial conditions. It cannot be broken down or decomposed into two subsystems (two invariant open subsets) which do not interact under $f$ because of topological transitivity. And, in the midst of this random behaviour, we nevertheless have an element of regularity, namely the periodic points which are dense'' \cite{Devaney89}. In other words, topological transitivity means that the dynamical system $f$ is, in a sense, indecomposable in simpler systems, and any given region of its phase space will eventually overlap with any other given region. This also means that, taken two points $P$ and $Q$, their trajectories, initially close together, can suddenly go in completely different directions.
An orbit (or a point or a trajectory) that repeats is called a periodic orbit. The density of periodic points implies that every point in the space is approached arbitrarily closely by periodic orbits. This excludes that a chaotic system can be in some way periodic and it also means that, taken two points $P$ and $Q$, their trajectories, initially far apart, can wind up in almost the same place.

In literature, it's widely used an ``experimental'' definition of chaos, that takes into account only the first condition of Devaney's definition, the sensitive dependence on initial conditions. There are a lot of numerical tests for this property, but this definition of chaos, as we already said, is not satisfactory. See, for instance, the counterexample 3.3 in \cite{Martelli98}.

Over the years, chaos theory gradually has provided a framework to study some interesting properties of time series. Some widespread tests are: correlation dimension, the BDS test, Kolmogrov entropy, Lyapunov exponent, close returns test, etc (see Section \ref{sec:exp} for a survey of these tests and a comparison with our approach). Some of them are not properly chaos tests but can be able to investigate properties like nonlinearity \footnote{Chaotic dynamics are necessarily non-linear, but there are many examples of non-linear dependencies that are not consistent with chaos, as obtained in \cite{Adrangi01}.}. As pointed out in \cite{yousefpoor2008looking}, some tests, which are not chaos tests in fact, have been implemented in chaos literature (especially in applied science such as, for our aims, in energy markets). A chaotic system must be characterized by some basic properties, while the tests mentioned above focus only on one of these aspects. Although this is not sufficient to ensure the presence of chaos, these tests could be useful to study important properties of system dynamics (such as butterfly effect).

%The celebrated butterfly effect is not equal to chaos, but, in order to be chaotic, it should satisfy other properties. However, an ``experimental'' definition of chaos is widely used by many non-mathematicians, particularly from physicists, engineers and economists, and it is based only on sensitive dependence on initial conditions. As we will discuss in Section \ref{sec:defchaos}, this definition of chaos is not satisfactory. Devaney's mathematical definition of chaos \cite{Devaney89} provides a useful framework for selecting chaos tests.

%Our paper is concerned about sensitive dependence on initial conditions test, its discussion in the overall framework of definition of chaos and, according to this framework, on rereading of existent works on energy markets.
%As our case study, some sample stocks were selected from \textbf{[QUALE MERCATO???]} and the chosen tests implemented on their returns time series.

Chaotic time paths have several properties that should be of special interest to commodity market observers, such as the apparent stochasticity of time series that could be generated by deterministic systems, or the butterfly effect.

In particular, since the energy futures time series show an irregular random behaviour that often resembles chaos, we employ the determinism test introduced by Kaplan and Glass \cite{Ka92}, in order to verify whether the studied time series indeed originates from a deterministic system. For this purpose, we use a package written in $C++$ code and developed in \cite{Ko05}. The importance of this procedure in the framework implemented here will be discussed in Section \ref{sec:exp}. The idea is to take the \emph{determinism coefficient} $\kappa$, obtained by Kaplan and Glass' test, as a measure of the reliability level (in percentage) of test on sensitive dependence on initial conditions. %In the light of the above,
Actually, it is known  that time series generated from stochastic systems also might show positive maximal Lyapunov exponents \cite{tanaka1996lyapunov}, \cite{ikeguchi1997lyapunov}, \cite{tanaka1998analysis}.

\begin{figure}[tb]
\centering
\includegraphics[scale=0.60]{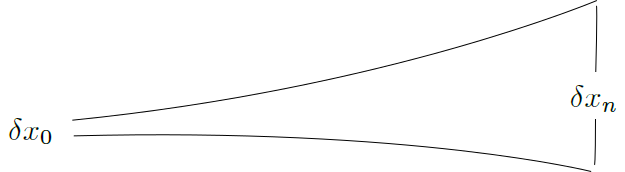}
\caption{Effect of a small change of initial condition.}
\label{fig:1}
\end{figure}

The butterfly effect has important implications in forecasting of chaotic time series.
Let us consider the time series $y_n$ and assume that there exists a system $(g,f, x_0)$ such that $y_n=g(x_n)$, $x_{n+1}=f(x_n)$, where $x_0$ is the initial condition at initial time $n=0$, and where $g$ maps the $m$-dimensional phase space $\mathbb R^m$ to $\mathbb R$, and $f$ maps $\mathbb R^m$ to $\mathbb R^m$. The function $f$ maps an unknown (to the econometrician) dynamic that governs the evolution of the unknown (to the econometrician) state $x_0$. The econometrician observes $y_n$. The task is to uncover information about $(g,f, x_0)$ from observations $y_n$ \cite{brock1988business}. The time series $y_n$, which we will assume as the data time series under analysis, has a chaotic explanation if $x_n$ is chaotic.
The question is whether it is possible to forecast a chaotic series. Intuitively, the butterfly effect usually doesn't allow long-term forecasting of chaotic series. If we change slightly the value of initial point: $x_0$ $\mapsto$ $x'_0=x_0+\delta x_0$, the point $x_n$ at discrete time $n$ will also be changed (see Fig. \ref{fig:1}).
What may happen is that, when time becomes large, the small initial distance $\delta x_0$ grows anyway, and it may grow exponentially fast: $\delta x_n\sim \delta x_0 e^{\lambda n}$, for some $\lambda>0$. The term $\delta x_n$ represents the uncertainty induced by perturbations. Hence, fixed $\delta x_0$ and $\lambda$, it is bounded if $e^{\lambda n}$ is bounded and so if $n$ is as small as possible ($n=1$ or $n=2$, for instance): this is the short-term forecasting of chaotic series, which is possible because $\delta x_n$ is amplified at a finite rate $e^{\lambda n}$. Accordingly, for chaotic time series, if one knows $(g,f)$ and could measure $x_n$ without error, one could forecast $x_{n+i}$ and, thus, $y_{n+i}$ perfectly in the short time (say a few days when dealing with daily data).

\section{The empirical results}
\label{sec:exp}

%In the Conclusions of \cite{matilla2007nonlinear}, Matilla-Garc\'{i}a asks: ``A natural question arises: Does evidence of chaos depend on the test procedure used by the researcher? This question is left for future research''. Obviously, the experimental results should be independent from the test used to obtain them but the basic question is the adequacy of the data used to test the chaotic structure.
%Chaos tests can be conducted on both the original as well as the filtered data. Empirical analysis of identification of chaotic structure of return series necessarily raises the question of filtering (the transformation of the original data prior to its analysis), because the filtered data may give a false indication of chaos. In \cite{Panas00}, authors compare filtered and raw data, pointing out that they are not very different among them, even if filtering may affect the dimensionality of the original data (see \cite{chen1993searching}, as reported also in \cite{Panas00}) and the filtered data may mimic a chaotic one. However, in the approach taken here we have used only the original data and we left the question of the filtering to future works.

In this Section we test for sensitive dependence on initial conditions and determinism of the following energy futures series: heating oil (06.03.1979 - 15.05.2014), natural gas (03.04.1990 - 15.05.2014). The time series were taken by NYMEX and were obtained from \url{http://www.quandl.com}. In particular, we investigate the presence of low-dimensional chaotic structure. Low-dimensional chaos is usually used to refer to dynamics with only one positive Lyapunov exponent \cite{harrison1999route}, while, on the contrary, high-dimensional chaos corresponds to more than one positive Lyapunov exponents. High and low-dimensional chaos in this sense are not linked to the embedding dimension. Although the treatment of this difference is beyond the scope of this paper and will be treated elsewhere, considering a three-dimensional system with the Lyapunov exponent spectrum $(\lambda_1,\lambda_2,\lambda_3)$, it has been shown that $(+, 0,-)$ indicates a chaotic attractor. Hence, if by definition, a high-dimensional chaotic system has at least two positive Lyapunov exponents, the Lyapunov exponent spectrum is $(\lambda_1,\lambda_2,\lambda_3,\lambda_4)=(+,+,0,-)$, and thus the minimal dimension for a high-dimensional chaotic system is $m=4$. This also means that, for $m<4$, we can have only the presence of low-dimensional chaos.
%a deterministic structure with a low embedding dimension

In order to investigate the presence of low-dimensional chaotic structure, fixing the embedding dimension $m=2$ or $m=3$, for each commodities we analyze the determinism rate $\kappa$ on varying of $\tau$. We use a package written in $C++$ code, which can be downloaded from M. Perc's Web page \footnote{M. Perc Web Page, \url{http://www.matjazperc.com/ejp/time.html}.}, as described in \cite{Ko05}.

To calculate the maximal Lyapunov exponent $\lambda_{max}$, we used a Matlab \cite{Matlab} code due to Wolf et al. and available on Matlab Central File Exchange. It is based on an algorithm presented in \cite{Wolf85}, that estimates the dominant Lyapunov exponent of a 1-D time series by monitoring orbital divergence. The algorithm was distributed for many years by the authors in Fortran and C and subsequently it has just been converted to Matlab. This algorithm has known a widespread success in literature: suffice it to say that while we write this paper, the work due to Wolf et al. was cited 6963 times (source: Google Scholar).

The main empirical results obtained in our analysis are summarised in Tables \ref{tab:tabella1a}, \ref{tab:tabella1b}, \ref{tab:tabella2a} and \ref{tab:tabella2b}. In them one uses results on determinism coefficient shown in Appendix. The Tables exhibit Maximum Lyapunov exponents for the cases with higher value determinism coefficient.
\begin{center}
\begin{tabular}{p{0.5cm}p{0.5cm}p{1.5cm}p{1.5cm}}
m &$\tau$&$\kappa$&MLE\\ \hline
2 &2& 0.501905 & 2.7159\\ \hline
2 &39&0.444272 & 1.7691 \\ \hline
2 &4& 0.441020 & 2.5150 \\ \hline
2 &28& 0.433819& 2.0525 \\ \hline
2 &5& 0.431399 & 2.4318 \\ \hline
\end{tabular}
\captionof{table}{Maximum Lyapunov exponent (MLE) for Heating OIL HO1, temporal range 06.03.1979 - 15.05.2014, number of samples 8.825. Case $m=2$.}
\label{tab:tabella1a}
\end{center}

\begin{center}
\begin{tabular}{p{0.5cm}p{0.5cm}p{1.5cm}p{1.5cm}}
m &$\tau$&$\kappa$&MLE\\ \hline
3 &2& 0.537274 & 2.2768  \\ \hline
3 &4& 0.485819 & 2.3791  \\ \hline
3 &3& 0.478729 & 2.6322  \\ \hline
3 &5& 0.469815 & 2.2892    \\ \hline
3 &7& 0.454737 & 2.0076  \\ \hline
\end{tabular}
\captionof{table}{Maximum Lyapunov exponent (MLE) for Heating OIL HO1, temporal range 06.03.1979 - 15.05.2014, number of samples 8.825. Case $m=3$.}
\label{tab:tabella1b}
\end{center}

\begin{center}
\begin{tabular}{p{0.5cm}p{0.5cm}p{1.5cm}p{1.5cm}}
m &$\tau$&$\kappa$&MLE\\ \hline
2 &16& 0.493609 & 12.5925\\ \hline
2 &14& 0.477181 & 12.3958 \\ \hline
2 &2& 0.476278 &  13.9755\\ \hline
2 &15& 0.459651 & 11.6615\\ \hline
2 &8& 0.452906 &  13.1308\\ \hline
\end{tabular}
\captionof{table}{Maximum Lyapunov exponent (MLE) for Natural GAS, temporal range 03.04.1990 - 15.05.2014, number of samples 6.042. Case $m=2$.}
\label{tab:tabella2a}
\end{center}

\begin{center}
\begin{tabular}{p{0.5cm}p{0.5cm}p{1.5cm}p{1.5cm}}
m &$\tau$&$\kappa$&MLE\\ \hline
3 &2& 0.565059 & 12.1163  \\ \hline
3 &3& 0.483286 & 11.8509  \\ \hline
3 &16& 0.476005 &8.9366  \\ \hline
3 &8& 0.473239 & 10.4472    \\ \hline
3 &4& 0.472915 & 10.9943  \\ \hline
\end{tabular}
\captionof{table}{Maximum Lyapunov exponent (MLE) for Natural GAS, temporal range 03.04.1990 - 15.05.2014, number of samples 6.042. Case $m=3$.}
\label{tab:tabella2b}
\end{center}

These results suggest that there is no strong evidence of sensitive dependence on initial conditions (butterfly effect) for the series considered here because $\kappa$ is not near to $1$.

To better understand the results obtained in our paper, let's take, for example, Table \ref{tab:tabella2a}. When $m=2$ and $\tau=16$ or $\tau=8$, natural gas time series exhibits positive maximal Lyapunov exponents with determinism coefficient, $\kappa=0.493609$ and $\kappa=0.452906$, respectively. Since there are stochastic systems that show sensitive dependence on initial conditions (\cite{tanaka1996lyapunov}, \cite{ikeguchi1997lyapunov}, \cite{tanaka1998analysis}), we propose to explain these results on $\kappa$ as the reliability level (in percentage) of the maximal Lyapunov exponents results. Thus, when $m=2$ and $\tau=16$ natural gas time series exhibits positive maximal Lyapunov exponents with reliability level at $\simeq 49\% $: too low to be able to conclude that there is strong evidence of sensitive dependence on initial conditions. %The adoption of a reliability level for the maximal Lyapunov exponents results is motivated by the fact that time series generated from stochastic systems also might show positive maximal Lyapunov exponents \cite{tanaka1996lyapunov}, \cite{ikeguchi1997lyapunov}, \cite{tanaka1998analysis}.

We recall that all the papers concerning energy time series that we have cited in this paper (\cite{chwee1998chaos}, \cite{serletis1999north}, \cite{Panas00}, \cite{Adrangi01}, \cite{moshiri2006forecasting}, \cite{matilla2007nonlinear}, \cite{kyrtsou2009energy}, \cite{barkoulas2012metric}) are based on the ``experimental'' definition of chaos. Among them, \cite{serletis1999north}, \cite{Panas00} and \cite{matilla2007nonlinear} discover a certain evidence of butterfly effect. Panas and Ninni \cite{Panas00} employ the BDS statistical test and the Correlation Dimension test to distinguish determinism by stochastic process; with the same aim, Matilla-Garc\'{i}a employs the BDS statistical test and the Kaplan test (see \cite{matilla2007nonlinear} for details) while Serletis and Gogas \cite{serletis1999north} apply a nonlinear analysis in order to remove any stochastic dependence.

With regard to the futures time series analyzed in our paper, we notice the following differences respect to other papers.
\begin{itemize}
  \item Kyrtsou et al. \cite{kyrtsou2009energy} have observed natural gas and heating oil, over the period from 1994 to mid-January 2008, showing that Lyapunov exponent estimates are negative. They also reveal the existence of a structure that is partially deterministic. The largest Lyapunov exponents detected in our paper for natural gas and heating oil are positive but the futures time series show a not negligible contribute of stochasticity ($\kappa\lesssim 50\% $ for all the measures).
  \item As for heating oil, Adrangi et al. \cite{Adrangi01}, for observations on the range 1/02/85 - 03/31/95, employ correlation dimension test, the BDS test and Kolmogrov entropy, without finding evidence of butterfly effect. In Table \ref{tab:tabella3}, we have calculated the \emph{correlation dimension} \footnote{File c2.m from D. Chelidze's home page, \url{http://egr.uri.edu/nld/software/}.} of NYMEX-heating oil daily series considered here. We notice an absence of saturation, that doesn't provide evidence of chaotic structure, and thus confirming the results by Adrangi et al. obtained with correlation dimension method (\cite{Adrangi01}, Table 3, p. 416).
  \item As for natural gas, Matilla-Garc\'{i}a \cite{matilla2007nonlinear} uses observations from 04/03/1990 to 10/19/2005. He discovers the positivity of the largest Lyapunov exponent while Chwee \cite{chwee1998chaos}, examining observations from April 1990 to September 1996, shows no evidence of butterfly effect from the estimation of the Lyapunov spectra. The range considered from Matilla-Garc\'{i}a is approximately the same we have considered here (with obviously few years less). Matilla-Garc\'{i}a, testing for the butterfly effect, employed a method (by Rosenstein et al. \cite{rosenstein1993practical}) belonging to the same family of the one used here (by Wolf et al. \cite{Wolf85}), and thus his results on positivity of MLE no way they could be according to ours. %\footnote{cioe i suoi risultati non potevano essere che in accordo con i nostri: stessa serie e stesso metodo}.
      Since it is possible that Rosenstein's and Wolf's algorithms find positive values for the Lyapunov exponent also for any pure random process, his results are accompanied with a test \cite{fernandez2005testing} which has a deterministic process as the null hypothesis, while the alternative hypothesis is that of a stochastic process. However, we point out that there are several systems that have both stochastic and deterministic components, \cite{Ka92}, and only a ``null hypothesis'' could be insufficient to describe these cases. For this reason we have introduced the parameter $\kappa$.
  \item Again, as for natural gas, Serletis and Gogas, examining monthly data from 1985:2 to 1996:12, ``test for positivity of the dominant Lyapunov exponent. Before conducting such a nonlinear analysis, the data were rendered stationary and appropriately ﬁltered, in order to remove any linear as well as nonlinear stochastic dependence''. They have found evidence of butterfly effect in all natural gas liquids markets. Compared to their paper, in our work we didn't apply any filtering to data.
\end{itemize}
Moreover, although these studies investigate the presence of both stochastic and determinist components in time series, they don't provide any estimate of determinism rate existing in the analyzed data.

\begin{center}
\begin{tabular}{p{0.5cm}p{1.5cm}p{1.5cm}p{1.5cm}p{1.5cm}}
$\tau$ &$m=5$&$m=10$&$m=15$&$m=20$\\ \hline
5  &-0.0841&-0.1377&-0.1840&-0.2123  \\ \hline
10 &-0.0845&-0.1406&-0.1914&-0.2259  \\ \hline
15 &-0.0848&-0.1427&-0.2004&-0.2355  \\ \hline
20 &-0.0852&-0.1452&-0.2076&-0.2428  \\ \hline
25 &-0.0855&-0.1487&-0.2123&-0.2469  \\ \hline
30 &-0.0860&-0.1523&-0.2163&-0.2507  \\ \hline
35 &-0.0869&-0.1569&-0.2203&-0.2649  \\ \hline
40 &-0.0878&-0.1614&-0.2242&-0.2691  \\ \hline
\end{tabular}
\captionof{table}{The table reports $\log_{10}C(\epsilon)$, where $C(\epsilon)$ is the correlation integral, for the daily prices of NYMEX heating oil (06.03.1979 - 15.05.2014).}
\label{tab:tabella3}
\end{center}

\section{Conclusions}
\label{sec:conc}
%According to Devaney's Definition, the ingredients of chaos are: sensitive dependence on initial conditions; topological transitivity; density of periodic points. The sensitive dependence on initial conditions is the most popular property of a chaotic system, maybe for its intuitive meaning: tiny differences become amplified. The shorthand is the ``butterfly effect'', a term that has inspired novels and movies. (Who remembers, for example, Ian Malcolm, the mathematician of Jurassic Park?)

%Nevertheless, as we have seen in Section \ref{sec:defchaos}, sensitive dependence on initial conditions is redundant with other two conditions of Devaney's Definition, even if an ``experimental'' definition of chaos (i.e., a definition which requires only the sensitive dependence) continues to enjoy considerable success between many non-mathematicians, e.g. in energy markets analysis.

%All of the results about energy markets cited in this paper (\cite{chwee1998chaos}, \cite{serletis1999north}, \cite{Panas00}, \cite{Adrangi01}, \cite{moshiri2006forecasting}, \cite{matilla2007nonlinear}, \cite{barkoulas2012metric} for the ``metric methodologies'') are concerned about an ``experimental'' definition of chaos. Despite it can be checked numerically, this definition of chaos is not satisfactory. See, for instance, the counterexample 3.3 in \cite{Martelli98}. The mathematical definition of chaos recalled here (in sense of Devany \cite{Devaney89}) is helpful to prevent us from misleading results about ostensible chaoticity of the returns series.

In this paper we have tested sensitive dependence on initial conditions and determinism of the NYMEX energy futures: heating oil (06.03.1979 - 15.05.2014), natural gas (03.04.1990 - 15.05.2014). Fixing the embedding dimension $m=2$ or $m=3$, for each commodities we have analyzed the determinism rate $\kappa$ on varying of $\tau$, employing a package developed in \cite{Ko05} and based on the determinism test introduced in \cite{Ka92}. Afterward, to check sensitive dependence on initial conditions we have selected the largest Lyapunov exponents test, using a Matlab code based on an algorithm presented in \cite{Wolf85}. The main empirical results obtained in the above steps are summarised in Tables \ref{tab:tabella1a}, \ref{tab:tabella1b}, \ref{tab:tabella2a} and \ref{tab:tabella2b}. Since time series generated from stochastic systems also might show positive maximal Lyapunov exponents \cite{tanaka1996lyapunov}, \cite{ikeguchi1997lyapunov}, \cite{tanaka1998analysis}, we have introduced a reliability level ($\kappa$, expressed in percentage) associated to maximal Lyapunov exponents results. The maximum reliability level obtained here was been $\simeq 56\% $, too low to be able to ensure that there is strong evidence of sensitive dependence on initial conditions.

Different approaches from those employed in this paper, such as the 0-1 test for chaos \cite{Gottwald04}-\cite{Gottwald05} or the recurrence plot analysis also used by Barkoulas et al. \cite{barkoulas2012metric}, deserve further discussions and they will be treated in the next works. Moreover, in future developments, it would be interesting to compare the results obtained in previous works employing the Shannon entropy \cite{benedetto2015maximum, benedetto2016predictability} (see also \cite{genccay2010crash,gradojevic2008overnight} for the investigation of the evolution of the aggregate market expectations) with the results obtainable through Lyapunov exponents and Kolmogorov entropy.

\section{Appendix 1: Mathematical methodologies of the tests}
\label{sec:prel}

The sensitive dependence on initial conditions can be checked by Kolmogrov entropy test and the Largest Lyapunov exponents.

The definition of Kolmogorov entropy (or Kolmogorov-Sinai entropy \cite{kolmogorov1958new}, \cite{sinaui1959concept}, or metric entropy) $K$ can be found, for example, in \cite{farmer1982chaotic} (eq. 7, p. 370). For systems that show butterfly effect, $K$ is positive\footnote{Farmer \cite{farmer1982chaotic}, Grassberger and Procaccia \cite{grassberger1983estimation} mention ``chaos'' but they refer to the experimental definition of chaos. In fact, Grassberger and Procaccia, defining the expression of $K$, cite the paper of Farmer, while Farmer defines ``chaotic attractor as any attractor with positive metric entropy'' (p. 372, \cite{farmer1982chaotic}) and, as we can read in the same page, $K$ can be expressed as the sum of positive Lyapunov exponents, which constitute a measure of butterfly effect and not chaos, as previously recalled.}. Grassberger and Procaccia \cite{grassberger1983estimation} have proposed a quantity $K_2$ which has the following properties: (i) $K_2 \geq0$; (ii) $K_2 \leq K$;
(iii) $K_2$ is infinite for random systems; and (iv) $K_2\neq0$ for systems which show butterfly effect. Nevertheless, the largest Lyapunov exponents test has been selected here, because of its more clear result. If the largest Lyapunov exponent is positive, then it would imply butterfly effect but if it is negative it would not, while $K_2 >0$ is only a sufficient condition for butterfly effect.

The existing determinism structure in the analyzed time series can be checked also by the correlation dimension test. The correlation dimension test measures a quantity called \emph{correlation dimension} and usually has been introduced as a test for distinguishing randomness and chaoticity. It distinguishes chaotic series from random series by investigating the correlation dimension behaviour of the data, \cite{gabisch2013business}.

Grassberger and Procaccia \cite{grassberger1983characterization} \cite{GRASSBERGER1983189} showed that correlation dimension $D_2$ can be evaluated using the correlation integral $C(\epsilon)$, which is defined to be the probability that a pair of points chosen randomly with respect to the natural measure is separated by a distance less than $\epsilon$ on the attractor. %For a trajectory of length $N$ in the embedding space $\mathbb R^m$, the correlation integral can be approximated by the sum:
%$$C_N(\epsilon)=\frac{2}{N(N-1)}\sum_{j=1}^N \sum_{i=j+1}^N\, \Theta\left( \epsilon - |\textbf{x}_i-\textbf{x}_j| \right),$$
%where $\Theta(\cdot)$ is the Heaviside function given by $\Theta(x)=1$ for $x\geq0$ and $0$ otherwise, and the norm is defined by $|\textbf{x}|=\max\{|x_i|: \, 1\leq i\leq m\}$.
%For $N$ large, we have $C_N(\epsilon)\approx C(\epsilon)$. Grassberger and Procaccia argue \footnote{dove hai preso questa roba?} that the correlation dimension is given by
%$$D_2=\lim_{\epsilon\rightarrow0}\lim_{N\rightarrow\infty}\, \frac{\ln C_N(\epsilon)}{\ln \epsilon}.$$
For a stochastic signal, $C(\epsilon)$ scales like $\epsilon^m$ for all $m$. In contrast, $C(\epsilon)$ scales like $\epsilon^{D_2}$ for $m$ larger than the attractor dimension, if the signal is generated by a deterministic system (\cite{GRASSBERGER1983189}, p. 206).

In other words, one can have the following cases: (i) if the value of $C(\epsilon)$ stabilizes at some value as $m$ increases, then the signal is generated by a deterministic system; (ii) if $C(\epsilon)$ continues to vary as $m$ is raised, then the system is to be regarded as stochastic. %, since for practical purposes, there is no difference between high-dimensional system and randomness \footnote{qualche referenza}.
When case (i) occurs, usually one tests for butterfly effect of the analyzed time series. Nevertheless, in this way, we don't have a measure of stochasticity (or determinism) rate of the time series. How reliable the results obtained on butterfly effect are?

In our paper Kaplan and Glass' algorithm (\cite{Ka92}, \cite{Ko05} and references therein) has been preferred over Grassberger and Procaccia algorithm, because of its more explicit result. As we have recalled, the determinism coefficient $\kappa$ is equal to 1 for a deterministic system, while for a random walk $\kappa=0$. In a way, $\kappa$ measures the distance of time series from a deterministic system and from a stochastic process. In other words, it measures both the rate of ``stochasticity '' and of ``determinism'' in a time series. While the Grassberger and Procaccia algorithm did not provide such explicit results. Hence, our idea is to take the $\kappa$ such as a measure of the reliability level (in percentage) of the results obtained with the test on butterfly effect.

For a deeper discussion of the mathematical aspects of the methodologies briefly described here, we refer to working paper \cite{MasVel}.

\section{Appendix 2}
\label{sec:app}
Fixed $m=2$ or $m=3$, in the following Tables we show the numerical values of determinism coefficient $\kappa$ on varying of $\tau$.

\setcounter{table}{0}
\renewcommand{\thetable}{A\arabic{table}}

{\fontsize{0.4cm}{0.3cm}\selectfont
\begin{center}
\begin{tabular}{p{1cm}p{1.5cm}}
$\tau$&$\kappa$\\ \hline
2&0.501905\\ \hline
3&0.419842\\ \hline
4&0.441020\\ \hline
5&0.431399\\ \hline
6&0.392880\\ \hline
7&0.395867\\ \hline
8&0.406614\\ \hline
9&0.415095\\ \hline
10&0.391700\\ \hline
11&0.415090\\ \hline
12&0.395824\\ \hline
13&0.375370\\ \hline
14&0.407810\\ \hline
15&0.408089\\ \hline
\end{tabular}
\begin{tabular}{p{1cm}p{1.5cm}}
$\tau$&$\kappa$\\ \hline
16&0.361204\\ \hline
17&0.375060\\ \hline
18&0.384556\\ \hline
19&0.398143\\ \hline
20&0.412437\\ \hline
21&0.406275\\ \hline
22&0.381689\\ \hline
23&0.395406\\ \hline
24&0.396505\\ \hline
25&0.409085\\ \hline
26&0.423537\\ \hline
27&0.419672\\ \hline
28&0.433819\\ \hline
29&0.428505\\ \hline
\end{tabular}
\begin{tabular}{p{1cm}p{1.5cm}}
$\tau$&$\kappa$\\ \hline
30&0.408889\\ \hline
31&0.411305\\ \hline
32&0.412225\\ \hline
33&0.379679\\ \hline
34&0.394530\\ \hline
35&0.394847\\ \hline
36&0.417282\\ \hline
37&0.430625\\ \hline
38&0.399006\\ \hline
39&0.444272\\ \hline
40&0.417778\\ \hline
41&0.384796\\ \hline
42&0.426004\\ \hline
43&0.395696\\ \hline
\end{tabular}
\captionof{table}{Heating oil, temporal range 06.03.1979 - 15.05.2014, number of samples 8.825, $m=2$.}
\label{tab:41}
\end{center}}

{\fontsize{0.4cm}{0.3cm}\selectfont
\begin{center}
\begin{tabular}{p{1cm}p{1.5cm}}
$\tau$&$\kappa$\\ \hline
2&0.537274\\ \hline
3&0.478729\\ \hline
4&0.485819\\ \hline
5&0.469815\\ \hline
6&0.438405\\ \hline
7&0.454737\\ \hline
8&0.420400\\ \hline
9&0.443673\\ \hline
10&0.406686\\ \hline
11&0.410421\\ \hline
12&0.384327\\ \hline
13&0.406481\\ \hline
14&0.431002\\ \hline
15&0.409526\\ \hline
\end{tabular}
\begin{tabular}{p{1cm}p{1.5cm}}
$\tau$&$\kappa$\\ \hline
16&0.394520\\ \hline
17&0.403045\\ \hline
18&0.401586\\ \hline
19&0.399474\\ \hline
20&0.414770\\ \hline
21&0.420266\\ \hline
22&0.404316\\ \hline
23&0.406564\\ \hline
24&0.384466\\ \hline
25&0.368297\\ \hline
26&0.393778\\ \hline
27&0.353556\\ \hline
28&0.398153\\ \hline
29&0.390039\\ \hline
\end{tabular}
\begin{tabular}{p{1cm}p{1.5cm}}
$\tau$&$\kappa$\\ \hline
30&0.367590\\ \hline
31&0.404781\\ \hline
32&0.415901\\ \hline
33&0.404190\\ \hline
34&0.412973\\ \hline
35&0.414922\\ \hline
36&0.438795\\ \hline
37&0.401609\\ \hline
38&0.414218\\ \hline
39&0.395303\\ \hline
40&0.422266\\ \hline
41&0.399929\\ \hline
42&0.381433\\ \hline
43&0.406102\\ \hline
\end{tabular}
\captionof{table}{Heating oil, temporal range 06.03.1979 - 15.05.2014, number of samples 8.825, $m=3$.}
\label{tab:42}
\end{center}}

{\fontsize{0.4cm}{0.3cm}\selectfont
\begin{center}
\begin{tabular}{p{1cm}p{1.5cm}}
$\tau$&$\kappa$\\ \hline
2&0.476278\\ \hline
3&0.423472\\ \hline
4&0.427403\\ \hline
5&0.393378\\ \hline
6&0.420013\\ \hline
7&0.451157\\ \hline
8&0.452906\\ \hline
9&0.436177\\ \hline
10&0.451256\\ \hline
11&0.431307\\ \hline
12&0.419933\\ \hline
13&0.419721\\ \hline
14&0.477181\\ \hline
15&0.459651\\ \hline
\end{tabular}
\begin{tabular}{p{1cm}p{1.5cm}}
$\tau$&$\kappa$\\ \hline
16&0.493609\\ \hline
17&0.432316\\ \hline
18&0.407716\\ \hline
19&0.417947\\ \hline
20&0.417331\\ \hline
21&0.437138\\ \hline
22&0.411759\\ \hline
23&0.402592\\ \hline
24&0.411774\\ \hline
25&0.418427\\ \hline
26&0.378412\\ \hline
27&0.412978\\ \hline
28&0.401562\\ \hline
29&0.433165\\ \hline
\end{tabular}
\begin{tabular}{p{1cm}p{1.5cm}}
$\tau$&$\kappa$\\ \hline
30&0.431342\\ \hline
31&0.419427\\ \hline
32&0.432981\\ \hline
33&0.427012\\ \hline
34&0.410806\\ \hline
35&0.429270\\ \hline
36&0.421174\\ \hline
37&0.446244\\ \hline
38&0.369850\\ \hline
39&0.402403\\ \hline
40&0.387175\\ \hline
41&0.440380\\ \hline
42&0.445409\\ \hline
43&0.426742\\ \hline
\end{tabular}
\captionof{table}{Natural gas, temporal range 03.04.1990 - 15.05.2014, number of samples 6.042, $m=2$.}
\label{tab:47}
\end{center}}

{\fontsize{0.4cm}{0.3cm}\selectfont
\begin{center}
\begin{tabular}{p{1cm}p{1.5cm}}
$\tau$&$\kappa$\\ \hline
2&0.565059\\ \hline
3&0.483286\\ \hline
4&0.472915\\ \hline
5&0.448891\\ \hline
6&0.420842\\ \hline
7&0.459661\\ \hline
8&0.473239\\ \hline
9&0.431479\\ \hline
10&0.439751\\ \hline
11&0.421026\\ \hline
12&0.403814\\ \hline
13&0.427620\\ \hline
14&0.446716\\ \hline
15&0.455369\\ \hline
\end{tabular}
\begin{tabular}{p{1cm}p{1.5cm}}
$\tau$&$\kappa$\\ \hline
16&0.476005\\ \hline
17&0.457383\\ \hline
18&0.418325\\ \hline
19&0.430515\\ \hline
20&0.433987\\ \hline
21&0.459805\\ \hline
22&0.422080\\ \hline
23&0.423422\\ \hline
24&0.411347\\ \hline
25&0.450880\\ \hline
26&0.418336\\ \hline
27&0.420452\\ \hline
28&0.421035\\ \hline
29&0.422340\\ \hline
\end{tabular}
\begin{tabular}{p{1cm}p{1.5cm}}
$\tau$&$\kappa$\\ \hline
30&0.448111\\ \hline
31&0.416498\\ \hline
32&0.426808\\ \hline
33&0.417782\\ \hline
34&0.411756\\ \hline
35&0.399251\\ \hline
36&0.396002\\ \hline
37&0.441582\\ \hline
38&0.387324\\ \hline
39&0.421272\\ \hline
40&0.393928\\ \hline
41&0.440154\\ \hline
42&0.460074\\ \hline
43&0.425142\\ \hline
\end{tabular}
\captionof{table}{Natural gas, temporal range 03.04.1990 - 15.05.2014, number of samples 6.042, $m=3$.}
\label{tab:48}
\end{center}}

\section*{Acknowledgements}

The authors thank Prof. Matja\u{z} Perc for the C++ code of the package developed in \cite{Ko05}.

%%%%%%%%%%%%%%%%%%%%%%%%%%%%%%%%%%%%%%%%%%%%%%%%%%%%%%%%%%%%%%%%%%%%%%

\section*{References}

\bibliography{energetic}

\end{document}